\documentclass[aps,prd,onecolumn,groupedaddress,showpacs,nofootinbib,amssymb]{revtex4}
\usepackage{graphicx,bm,color}
\usepackage{amsmath}
\usepackage{amssymb}
\usepackage{amsfonts}

\newcommand{\be}{\begin{equation}}
\newcommand{\ee}{\end{equation}}
\newcommand{\bea}{\begin{eqnarray}}
\newcommand{\eea}{\end{eqnarray}}
\newcommand{\beaa}{\begin{eqnarray*}}
\newcommand{\eeaa}{\end{eqnarray*}}

\newcommand{\nn}{\nonumber \\}
\newcommand{\e}{\mathrm{e}}

\allowdisplaybreaks[4]

\begin{document}

\tolerance=5000

\title{Instabilities and Anti-Evaporation of Reissner-Nordstr\"{o}m Black Holes in
modified $F(R)$ gravity }

\author{Shin'ichi Nojiri$^{1,2}$ and Sergei D. Odintsov$^{3,4,5}$}

\affiliation{ $^1$ Department of Physics, Nagoya University, Nagoya
464-8602, Japan \\
$^2$ Kobayashi-Maskawa Institute for the Origin of Particles and
the Universe, Nagoya University, Nagoya 464-8602, Japan \\
$^3$Instituci\`{o} Catalana de Recerca i Estudis Avan\c{c}ats
(ICREA) and Institut de Ciencies de l'Espai (IEEC-CSIC), C5, Campus UAB, 08193
Bellaterra (Barcelona), Spain \\
$^4$  Tomsk State Pedagogical University, 
634061 Tomsk and National Research Tomsk State University, 634050 Tomsk, 
Russia \\
$^5$ King Abdulaziz University, Jeddah, Saudi Arabia}

\begin{abstract}
We study the instabilities and related anti-evaporation of the extremal
Reissner-Nordstr\"{o}m (RN) black hole in $F(R)$ gravity.
It is remarkable that  the effective electric charge can be generated for 
some solutions of  $F(R)$ gravity without electromagnetic field. The
anti-evaporation effect occurs but it  emerges
only in the strong coupling limit of the effective gravitational coupling.
The instabilities of RN black hole are also investigated when  the
electromagnetic sector is added to the action of
$F(R)$ gravity. We show the anti-evaporation occurs  in the Maxwell-$F(R)$
gravity  with the arbitrary gravitational coupling
constant although  it does not occur in the Maxwell-Einstein
gravity. Furthermore, general spherically-symmetric solution of $F(R)$
gravity in the Einstein frame is obtained.

\end{abstract}


\maketitle

\section{Introduction}

About forty years ago, Hawking demonstrated that black holes lose their energy
by the radiation generated by quantum effects \cite{Hawking:1974sw}.
As a result, the horizon radius of the black hole in the vacuum usually
decreases.
Bousso and Hawking \cite{Bousso:1997wi}, however,  have suggested a phenomenon
where the black hole radius increases by the quantum correction for the specific
(degenerate) Nariai black hole \cite{Nariai}. Such effect  is named as 
black hole anti-evaporation.
In the Schwarzschild- de-Sitter space-time, there are two horizons, that is,
the black hole horizon and the cosmological horizon.
The Nariai space-time can be obtained by the limit where the radius of the
cosmological horizon coincides with the radius of the black hole horizon.
It is  proposed that such black holes do not appear by the
collapse of stars but they could be primordial ones.
Usually, such primordial black holes should quickly evaporate by the Hawking
radiation because the smaller black holes have higher temperatures.
Therefore such primordial black holes have not been expected to be observed in
the present universe.
The first discussion \cite{Bousso:1997wi}  about the anti-evaporation is based
on the account of the quantum effects in $s$-wave and one-loop approximation,
where four-dimensional space-time is effectively reduced to two-dimensional space
with dynamical dilaton. (For more general study in the same approximation, 
see Ref.~\cite{bou} where instabilities of such black holes with 
evaporation/anti-evaporation regimes have been investigated.)
Using more general two-dimensional effective action more complete 
arguments were given
in \cite{Nojiri:1998ue} and more consistent four-dimensional approach with account
of the full four-dimensional conformal anomaly induced effective action was given in
Ref.~\cite{Nojiri:1998ph} because  $s$-wave approximation may be often not
reliable.
It has been proved that the anti-evaporation could occur in the full
four-dimensional quantum theory.

Recently in \cite{Nojiri:2013su}, it has been shown that the anti-evaporation
of the Nariai space-time may occur  already on classical level in modified $F(R)$
gravity. Note that such theory (for extended review, see \cite{review}) has
higher-derivative structure and in many cases may be considered as approximated
effective action for quantum gravity.
In this paper, we investigate the instabilities and anti-evaporation of charged
black  holes in $F(R)$ gravity.
In $F(R)$ gravity, there exist solutions which describe the
RN space-time even if there are no the Maxwell fields
\cite{Hendi:2009sw}.
As well-known, charged black hole which is called  the
RN space-time, has two horizons.
For some versions of $F(R)$ gravity, the RN space-time is
a solution even if there are no  Maxwell fields.
In this paper (Sect.II), we present the general conditions for the form of
$F(R)$ so that there exist solutions corresponding to  RN space-time.
In section III we consider the extremal limit where the radiuses of the
horizons coincide with each
other. There appears a space-time, which is similar to the Nariai 
space-time but
the signature of  the metric is different.
In case of the usual Einstein gravity or gravity with dilaton, the
anti-evaporation of such extremal solution has been investigated in
\cite{Nojiri:1998yg}.
Working with extremal RN black hole in $F(R)$ gravity, we show that there could
occur the anti-evaporation but only in the region with very strong
gravitational coupling. In section IV the same anti-evaporation effect is
studied in Maxwell-$F(R)$ gravity. In this case, its occurrence seems to be
quite realistic while the effect itself is fully attributed to $F(R)$
gravitational contribution.
Some summary and outlook is given in the Discussion.
Useful expressions for curvature are given in Appendix A.
In the Appendix B we construct general spherically-symmetric solution of $F(R)$
gravity in Einstein frame. Such solution is starting point for discussion
of the instabilities of black holes in this frame.

\section{Charged black hole solution in $F(R)$ gravity}

It is rather trivial to show the GR vacuum solutions like the Schwarzschild or
the Kerr-(anti-) de Sitter space-time  are also vacuum solutions in $F(R)$
gravity. It is known that RN space-time can be a vacuum solution
for a class of $F(R)$ gravity 
(for different black hole solutions in $F(R)$ gravity, 
see Refs.~\cite{brevik,Hendi:2012nj,A,B}.).
In the Einstein gravity, the RN solution only appears when the Maxwell action is
taken into account. In case of $F(R)$ gravity, the RN space-time
can be generated without the Maxwell sector.
In this section, we show the general conditions under which the RN space-time
becomes a solution of $F(R)$ gravity.

We now consider the following spherically symmetric and static metric,
\be
\label{BH1}
ds^2 = - A(r) dt^2 + B(r) dr^2 + r^2 d\Omega^2\, .
\ee
Here $\Omega^2$ is the metric of two-dimensional unit sphere.
We further assume
\be
\label{BH2}
B=\frac{1}{A}\, ,
\ee
and the scalar curvature is constant:
\be
\label{BH3}
R=R_0\, .
\ee
Then we obtain the following equation:
\be
\label{BH4}
\frac{d^2 A}{dr^2} + \frac{4}{r}\frac{d A}{dr} + \frac{2}{r^2}
= \frac{2}{r^2} - R_0\, ,
\ee
whose general solution is given by
\be
\label{BH5}
A = 1 - \frac{R_0 r^2}{12} - \frac{M}{r} + \frac{Q}{r^2}\, ,
\ee
which expresses nothing but the Reissner-Nordstr\''{o}m-(anti)-de Sitter
space-time.

General field equation in $F(R)$ gravity is given by
\be
\label{BH6}
0 = \frac{1}{2} g_{\mu\nu} F(R) - R_{\mu\nu} F'(R) + \nabla_\mu \nabla_\nu
F'(R) - g_{\mu\nu} \Box F'(R) + \frac{1}{2} T_{\mu\nu}\, .
\ee
Here $T_{\mu\nu}$ is the energy-momentum tensor of the matter field.
Now we neglect the contribution from the matter $\left( T_{\mu\nu}=0\right)$
and we assume the scalar curvature is constant (\ref{BH3}).
Then Eq.~(\ref{BH6}) reduces to
\be
\label{BH7}
0 = \frac{1}{2} g_{\mu\nu} F(R) - R_{\mu\nu} F'(R) \, .
\ee
By multiplying (\ref{BH7}) with $g^{\mu\nu}$, we obtain
\be
\label{BH8}
0 = 2 F(R) - R F'(R) \, .
\ee
By using (\ref{BH8}), one may eliminate $F(R)$ in (\ref{BH7}) and obtain
\be
\label{BH9}
0= \left( \frac{1}{4} g_{\mu\nu} R - R_{\mu\nu} \right) F'(R) \, .
\ee
By substituting the expression (\ref{BH5}), one obtains
\be
\label{BH10}
0=Q F'(R_0)\, .
\ee
Then, $Q=0$ or $F'(R_0)=0$.
The case $Q=0$  corresponds to the Schwarzschild-(anti)-de Sitter
space-time.
We may obtain the Reissner-Nordstr\''{o}m-(anti)-de Sitter (RNAdS) space-time
with $Q\neq 0$ if $F'(R_0)=0$, which tells $F(R_0)=0$ by using (\ref{BH8}).
Therefore the condition that the RNAdS space-time
is the solution is given by the existence of the scalar curvature $R=R_0$,
which satisfies
\be
\label{BH11}
0=F(R_0)=F'(R_0)\, .
\ee
A simple example is given by
\be
\label{BH12}
F(R) = \alpha \left( R - R_0 \right)^n\, ,\quad n\geq 2\, .
\ee
Here $\alpha$ is a constant.
The case of $R_0=0$ has been investigated in \cite{Hendi:2009sw}.

Thus, we have found the condition under which the
Reissner-Nordstr\''{o}m-(anti)-de
Sitter space-time becomes a solution of $F(R)$ gravity is given by  (\ref{BH11}).
In the Einstein gravity, $Q$ in (\ref{BH5}) corresponds to the electric charge of
the black hole. In $F(R)$ gravity which we are considering, there is no Maxwell
field but the charge can be effectively generated by $F(R)$. 
In (\ref{BH5}), the parameter $M$ could surely correspond to the mass of the black hole. 
The meaning of the parameter $Q$ is, however, not clear. 
The quantity $T^\mathrm{eff}_{\mu\nu} = R_{\mu\nu} - \frac{1}{4} g_{\mu\nu} R_0$ may 
be regarded as an effective energy-momentum tensor. 
Because $g^{\mu\nu} T^\mathrm{eff}_{\mu\nu} = 0$, that is, the trace of 
$T^\mathrm{eff}_{\mu\nu}$ vanishes, this effective energy-momentum tensor 
predends to be the energy-momentum tensor corresponding to the conformal 
or massless field like the Maxwell field. 
Some remark is in order.
It is known that $F(R)$ gravity can be rewritten in the scalar-tensor form,
where the scalar field corresponds to the scalar curvature.
Then someone might think that the charge $Q$ could correspond to the
scalar field but it is not true because we are considering the case of the
constant scalar curvature and therefore the scalar field is also constant. 
Then the scalar field cannot play the role of hair.

\section{Anti-evaporation of RN black hole in $F(R)$ gravity}

In this section, we consider the limit that the radiuses of two horizons
coincide with each other so that instability may occur.

In (\ref{BH5}), we may write $M$ and $Q$ in terms of two parameters $r_0$ and
$r_1$ as follows,
\be
\label{BH13}
M = \left( r_0 + r_1 \right) \left( 1 - \frac{\left(r_0^2 + r_1^2 \right)
R_0}{12} \right)\, ,
\quad
Q = r_0 r_1 \left( 1 - \frac{\left(r_0^2 + r_1^2 + r_0 r_1 \right) R_0}{12}
\right)\, .
\ee
Then we find
\be
\label{BH14}
A = \left( 1 - \frac{r_0}{r} \right) \left( 1 - \frac{r_1}{r} \right)
\left\{ 1 - \frac{ \left( \left( r + r_0 \right) \left( r + r_1 \right) + r_0^2
+ r_1^2 \right) R_0}{12} \right\}\, .
\ee
Therefore there are horizons at $r=r_0, r_1$.
When $R_0>0$ there is also a cosmological horizon.

We now rewrite the parameters and coordinates as follows,
\be
\label{BH15}
r_1 = r_0 + \epsilon \, , \quad r = r_0 + \frac{\epsilon}{2} \left( 1 + \tanh x
\right) \, ,
\ee
and consider the limit of $\epsilon\to 0$.
Then
\be
\label{BH16}
A \to - \frac{\epsilon^2}{4 r_0^2} \left( 1 - \frac{r_0^2 R_0}{2} \right)
\cosh^2 x\, .
\ee
By further redefining $t$ as
\be
\label{BH17}
t = \frac{2 r_0^2}{\epsilon \left( 1 - \frac{r_0^2 R_0}{2}\right)} \tau\, ,
\ee
we obtain the following metric
\be
\label{BH18}
ds^2 = \frac{r_0^2}{ \left( 1 - \frac{r_0^2 R_0}{2}\right)\cosh^2 x}
\left( d\tau^2 - dx^2 \right)  + r_0^2 d\Omega^2\, ,
\ee
which is similar to the metric of the Nariai space-time \cite{Nariai}:
\be
\label{Nr1}
ds_{\mathrm{Nariai}}^2 = \frac{1}{\Lambda^2 \cosh^2 x} \left( - d\tau^2 + dx^2
\right)
+ \frac{1}{\Lambda^2}d\Omega^2\, .
\ee
Here $\Lambda$ is a mass scale.
The signature for $\tau$ and $x$ are different from those in (\ref{BH18}).
If we neglect the signatures, when $R_0$ vanishes, we can identify $r_0= 1/\Lambda$.

We now consider the perturbation from the solution describing the space-time
similar to the Nariai space-time.
The metric is now assumed to be
\be
\label{Nr3}
ds^2 = \frac{\e^{2\rho\left(x,\tau\right)}}{\Lambda^2} \left( d\tau^2 - dx^2
\right)
+ \frac{\e^{-2 \varphi\left(x,\tau\right)}}{{\Lambda'}^2} d\Omega^2\, .
\ee
Here
\be
\label{BH19}
\Lambda =  \frac{ \sqrt{ 1 - \frac{r_0^2 R_0}{2}}}{r_0}\, , \quad
\Lambda' = \frac{1}{r_0}\, .
\ee
Then the $(\tau,\tau)$, $(x,x)$, $(\tau,x)$
$\left( \left(x,\tau\right) \right)$, and
$(\theta,\theta)$ $\left(\left(\phi,\phi\right)\right)$
components in (\ref{BH6}) have the following forms:
\begin{align}
\label{Nr5}
0 = & \frac{\e^{2\rho}}{2\Lambda^2} F(R) - \left( - \ddot\rho + 2 \ddot\varphi
+ \rho'' - 2 {\dot\varphi}^2 - 2 \rho' \varphi' - 2 \dot\rho \dot\varphi
\right) F'(R)
+ \frac{\partial^2 F'(R)}{\partial \tau^2} - \dot\rho \frac{\partial
F'(R)}{\partial \tau} - \rho' \frac{\partial F'(R)}{\partial x} \nn
& + \e^{2 \varphi} \left\{ - \frac{\partial}{\partial \tau}
\left( \e^{-2\varphi} \frac{\partial F'(R)}{\partial \tau}\right)
+ \frac{\partial}{\partial x}
\left( \e^{-2\varphi} \frac{\partial F'(R)}{\partial x}\right)\right\} \, ,\nn
0 = & - \frac{\e^{2\rho}}{2\Lambda^2} F(R) - \left( \ddot\rho + 2 \varphi''
   - \rho'' - 2 {\varphi'}^2 - 2 \rho' \varphi' - 2 \dot\rho \dot\varphi \right)
F'(R) + \frac{\partial^2 F'(R)}{\partial x^2} - \dot\rho \frac{\partial
F'(R)}{\partial \tau} - \rho' \frac{\partial F'(R)}{\partial x} \nn
& - \e^{2 \varphi} \left\{ - \frac{\partial}{\partial \tau}
\left( \e^{-2\varphi} \frac{\partial F'(R)}{\partial \tau}\right)
+ \frac{\partial}{\partial x}
\left( \e^{-2\varphi} \frac{\partial F'(R)}{\partial x}\right)\right\} \, ,\nn
0 = & - \left( 2 {\dot\varphi}' - 2 \varphi' \dot\varphi - 2\rho' \dot\varphi
   -2 \dot\rho \varphi' \right) F'(R)
+ \frac{\partial^2 F'(R)}{\partial \tau \partial x} - \dot\rho \frac{\partial
F'(R)}{\partial x} - \rho' \frac{\partial F'(R)}{\partial \tau} \, ,\nn
0 = & - \frac{\e^{- 2\varphi}}{2{\Lambda'}^2} F(R)
   - \frac{\Lambda^2}{{\Lambda'}^2}\e^{-2 \left(\rho+\varphi\right)}
\left( - \ddot\varphi + \varphi'' - 2 {\varphi'}^2 + 2 {\dot\varphi}^2 \right)
F'(R) + F'(R) \nn
& +  \frac{\Lambda^2}{{\Lambda'}^2}\e^{-2 \left(\rho+\varphi\right)} \left(
\dot\varphi \frac{\partial F'(R)}{\partial t}
    - \varphi' \frac{\partial F'(R)}{\partial x} \right)\nn
& - \frac{\Lambda^2}{{\Lambda'}^2} \e^{- 2\rho} \left\{ -
\frac{\partial}{\partial t}
\left( \e^{-2\varphi} \frac{\partial F'(R)}{\partial t}\right)
+ \frac{\partial}{\partial x}
\left( \e^{-2\varphi} \frac{\partial F'(R)}{\partial x}\right)\right\} \, .
\end{align}
Let us study  the following perturbation from the space-time similar to the
Nariai space-time:
\be
\label{Nr6}
\rho = - \ln \left( \cosh x \right) + \delta\rho\, ,\quad
\varphi = \delta \varphi\, .
\ee
Then one finds
\be
\label{Nr7}
\delta R = - 4 \Lambda^2 \delta\rho + 4 {\Lambda'}^2 \delta\varphi
   - \Lambda^2 \cosh^2 x \left( 2 \left(\delta \ddot\rho - \delta \rho'' \right)
   - 4 \left(\delta \ddot\varphi - \delta \varphi''\right) \right) \, ,
\ee
and the perturbed equations are
\begin{align}
\label{Nr8}
0 = & F'' \left( R_0 \right) \left\{- \frac{1}{ \cosh^2 x} \delta R
+ \tanh x \delta R' + \delta R'' \right\} \, ,\nn
0 = & F'' \left( R_0 \right) \left\{ \frac{1}{\cosh^2 x} \delta R
+ \delta \ddot R + \tanh x \delta R' \right\} \, ,\nn
0 = & F'' \left( R_0 \right) \left\{ \delta {\dot R}' + \tanh x \delta \dot R
\right\} \, ,\nn
0 = & F'' \left( R_0 \right) \left\{ \delta R - \cosh^2 x \left( - \delta \ddot
R + \delta R''
\right) \right\} \, .
\end{align}
Here we have used (\ref{BH11}).
If $F'' \left( R_0 \right) =0$ in addition to (\ref{BH11}), $\delta\rho$ and
$\delta\varphi$ can be arbitrary.
When  $F'' \left( R_0 \right) \neq 0$, by combining the first, second,
and fourth equations in (\ref{Nr8}), we obtain $\delta R=0$.

By using Eq.~(\ref{Nr7}), we find that the equation $\delta R=0$ has many
solutions.
By following \cite{Bousso:1997wi,Nojiri:1998ue,Nojiri:1998ph},
we may assume
\be
\label{BH20}
\delta\rho = \rho_0 \cosh \omega \tau \cosh^\beta x \, , \quad
\delta\varphi = \varphi_0 \cosh \omega \tau \cosh^\beta x\, ,
\ee
with constants $\rho_0$, $\varphi_0$, $\omega$, and $\beta$.
Then by substituting the expressions in (\ref{BH20}) into Eq.~(\ref{Nr7}) with
$\delta R=0$,
one gets
\be
\label{BH21}
0 = 2 {\Lambda'}^2 \left( \beta^2 - \beta + 1 \right) \varphi_0 - \Lambda^2
\left( \beta^2 - \beta + 4 \right) \rho_0\, , \quad
0= \left(\Lambda^2 \rho_0 - 2 {\Lambda'}^2 \varphi_0 \right) \left( \omega^2 -
\beta^2 \right)\, .
\ee
If $\Lambda^2 \rho_0 - 2 {\Lambda'}^2 \varphi_0 = 0$, the first equation in
(\ref{BH21}) tells $\rho_0 = \varphi_0 =0$.
Then we obtain
\be
\label{BH22}
\omega = \pm \beta\, .
\ee
For arbitrary $\beta$ and $\varphi_0$, there is a solution for $\rho_0$.
Therefore  $\beta$ and $\varphi_0$ can be arbitrary.

The horizon can be defined by
\be
\label{Nr18}
g^{\mu\nu}\nabla_\mu \varphi \nabla_\nu\varphi = 0\, .
\ee
By using (\ref{BH20}), we obtain
\be
\label{Nr19}
\tanh^2 \omega t = \tanh^2 x\, .
\ee
Here Eq.~(\ref{BH22}) is used.
Then we find \be
\label{Nr20}
\delta\varphi = \delta\varphi_\mathrm{h} \equiv
\varphi_0 \cosh^2 \beta t \, .
\ee
Eq.~(\ref{Nr3}) shows that $\e^{-\varphi}$ can be regarded as a radius
coordinate.
Using (\ref{Nr3}), one may now define the radius of the horizon by
$r_\mathrm{h} = \e^{-\delta\varphi_\mathrm{h}}/\Lambda$, then we find
\be
\label{Nr21}
r_\mathrm{h} = \frac{\e^{-\varphi_0 \cosh^2 \beta t}}{\Lambda}\, .
\ee
If $\varphi_0 < 0$, $r_\mathrm{h}$ increases,the appearing instability
may be regarded as anti-evaporation. 
We should note that the anti-evaporation occurs due to purely classical effect but 
if we include the quantum effects, there could also exist the usual Hawking radiation and 
evaporation. Then when we need qualitative arguments, we need to compare the strengths of 
the classical and quantum effects. 

Finally in this section, we reconsider the conditions (\ref{BH11}).
In $F(R)$ gravity, the effective gravitational coupling $\kappa_{\mathrm{eff}}$
is presented as $2 \kappa_\mathrm{eff}^2 = 1/F' \left(R_0\right)$.
Then the condition $F'(R_0)=0$ in (\ref{BH11}) tells that RN space-time occurs
only in the regime where the gravitational coupling is very strong. Clearly,
this is not very realistic situation.

We may consider the following model:
\be
\label{NN1}
F(R) = \frac{R}{2\kappa^2} \left( 1 - \frac{R}{R_0} \right)^2\, ,
\ee
which satisfies the conditions  (\ref{BH11}) at $R=R_0$.
When $R\to 0$, this model goes to the Einstein gravity, $F(R)\to R/2\kappa^2$.
Because
\be
\label{NN2}
F'(R) = \frac{1}{2\kappa^2} \left( 1 - \frac{R}{R_0} \right)
\left( 1 - \frac{3R}{R_0} \right)\, ,
\ee
there appears an anti-gravity region $R_0/3 < R < R_0$, that is,
the anti-gravity region lies between the Einstein gravity limit $R\to 0$ and
the point $R=R_0$ in  (\ref{BH11}) \cite{C}.
(For recent discussion of anti-gravity in $F(R)$
theory, see \cite{bamba}). This structure is rather general as long as $F(R)$
is the continuous function of $R$.
This indicates that the black hole with effective charge in (\ref{BH5}) cannot
be generated by the homogeneous evolution of the universe.
In the early universe, there could be the fluctuations as observed in BICEP2.
These fluctuations may create small domains where $R\sim R_0$ and generate
the RN black holes with effective charges.
Such primordial black holes may survive until now due
to the anti-evaporation.

\section{Charged black hole in $F(R)$ gravity}

As shown in the previous section, the condition to obtain the black hole
solution with effectively
generated charge in pure $F(R)$ theory, is the divergence of the effective
gravitational coupling $2 \kappa_\mathrm{eff}^2 = 1/F' \left(R_0\right)$.
In order to avoid this problem, we include real electromagnetic field $A_\mu$
and we start with the following action:
\be
\label{E0}
S = \int d^4 x \sqrt{ - g} \left\{ F(R) - \frac{1}{4} F^{\mu\nu} F_{\mu\nu}
\right\}\, .
\ee
Here $F_{\mu\nu}=\partial_\mu A_\nu - \partial_\nu A_\mu$.

In the spherically symmetric and static metric (\ref{BH1}) with (\ref{BH2}),
the equation for the vector field $A_\mu$ has the following form:
\be
\label{E1}
\frac{d}{dr} \left( r^2 F_{r0}\right)=0\, .
\ee
We now assume $A_i=0$ and $A_0$ depends only on the radial coordinate $r$.
Then (\ref{E1}) can be solved easily and we obtain
\be
\label{E2}
F_{rt} = \frac{q}{r^2}\, .
\ee
Here $q$ corresponds to the electric charge.
Then the stress-energy tensor of the vector field is given by
\be
\label{E3}
T_{tt} = - \frac{q^2 A}{2r^4}\, ,\quad T_{rr}= \frac{q^2}{2 r^4 A}\, ,\quad
T_{ij} = - \frac{q^2}{2r^2} {\tilde g}_{ij}\, .
\ee
Here we write $d\Omega^2 = \sum_{i,j=1,2}{\tilde g}_{ij} dx^i dx^j$.
Then in case that the scalar curvature is a constant (\ref{BH3}),
that is in case of (\ref{BH5}), by using general field equation in $F(R)$
gravity (\ref{BH6}), we find
\be
\label{E4}
F'\left(R_0\right) Q = q^2\, .
\ee
We now consider the perturbation from the solution describing the space-time
similar to the Nariai space-time.
The metric is now assumed as in (\ref{Nr3}).
In this space-time (\ref{Nr3}), we find that the electric field is given by
\be
\label{E5}
F_{x\tau}= \frac{q \e^{2\rho}}{\Lambda^2}\, .
\ee
Then the $(\tau,\tau)$, $(x,x)$, $(\tau,x)$
$\left( \left(x,\tau\right) \right)$, and
$(\theta,\theta)$ $\left(\left(\phi,\phi\right)\right)$
components in (\ref{BH6}) have the following forms:
\begin{align}
\label{Nr5B}
0 = & \frac{\e^{2\rho}}{2\Lambda^2} F(R) - \left( - \ddot\rho + 2 \ddot\varphi
+ \rho'' - 2 {\dot\varphi}^2 - 2 \rho' \varphi' - 2 \dot\rho \dot\varphi
\right) F'(R) + \frac{\partial^2 F'(R)}{\partial \tau^2} - \dot\rho \frac{\partial
F'(R)}{\partial \tau} - \rho' \frac{\partial F'(R)}{\partial x} \nn
& + \e^{2 \varphi} \left\{ - \frac{\partial}{\partial \tau}
\left( \e^{-2\varphi} \frac{\partial F'(R)}{\partial \tau}\right)
+ \frac{\partial}{\partial x}
\left( \e^{-2\varphi} \frac{\partial F'(R)}{\partial x}\right)\right\}
+ \frac{q^2 \e^{2\rho}}{2\Lambda^2} \, ,\nn
0 = & - \frac{\e^{2\rho}}{2\Lambda^2} F(R) - \left( \ddot\rho + 2 \varphi''
   - \rho'' - 2 {\varphi'}^2 - 2 \rho' \varphi' - 2 \dot\rho \dot\varphi \right)
F'(R) + \frac{\partial^2 F'(R)}{\partial x^2} - \dot\rho \frac{\partial
F'(R)}{\partial \tau} - \rho' \frac{\partial F'(R)}{\partial x} \nn
& - \e^{2 \varphi} \left\{ - \frac{\partial}{\partial \tau}
\left( \e^{-2\varphi} \frac{\partial F'(R)}{\partial \tau}\right)
+ \frac{\partial}{\partial x}
\left( \e^{-2\varphi} \frac{\partial F'(R)}{\partial x}\right)\right\}
- \frac{q^2 \e^{2\rho}}{2\Lambda^2} \, ,\nn
0 = & - \left( 2 {\dot\varphi}' - 2 \varphi' \dot\varphi - 2\rho' \dot\varphi
   -2 \dot\rho \varphi' \right) F'(R)
+ \frac{\partial^2 F'(R)}{\partial \tau \partial x} - \dot\rho \frac{\partial
F'(R)}{\partial x} - \rho' \frac{\partial F'(R)}{\partial \tau} \, ,\nn
0 = & - \frac{\e^{- 2\varphi}}{2{\Lambda'}^2} F(R)
   - \frac{\Lambda^2}{{\Lambda'}^2}\e^{-2 \left(\rho+\varphi\right)}
\left( - \ddot\varphi + \varphi'' - 2 {\varphi'}^2 + 2 {\dot\varphi}^2 \right)
F'(R) + F'(R) \nn
& +  \frac{\Lambda^2}{{\Lambda'}^2}\e^{-2 \left(\rho+\varphi\right)} \left(
\dot\varphi \frac{\partial F'(R)}{\partial t}
    - \varphi' \frac{\partial F'(R)}{\partial x} \right)\nn
& - \frac{\Lambda^2}{{\Lambda'}^2} \e^{- 2\rho} \left\{ -
\frac{\partial}{\partial t}
\left( \e^{-2\varphi} \frac{\partial F'(R)}{\partial t}\right)
+ \frac{\partial}{\partial x}
\left( \e^{-2\varphi} \frac{\partial F'(R)}{\partial x}\right)\right\}
+ \frac{q^2 \e^{-2\phi}}{2{\Lambda'}^2} \, .
\end{align}
Considering the perturbation in (\ref{Nr6}), we find
\begin{align}
\label{Nr8A}
0 = & \frac{ F'\left( R_0 \right) + 2 \Lambda^2 F'' \left( R_0 \right)}{2}
\delta R
   - F' \left( R_0 \right) \Lambda^2 \cosh^2 x
\left( - \delta \ddot\rho + 2 \delta \ddot\varphi
+ \delta \rho'' + 2 \tanh x \delta\varphi' \right) \nn
&  - 2 F' \Lambda^2 \left( R_0 \right) \delta\rho
+ F'' \left( R_0 \right) \Lambda^2 \cosh^2 x \left(
\tanh x \delta R' + \delta R'' \right) \, ,\\
\label{Nr8B}
0 = & - \frac{ F'\left( R_0 \right) + 2 \Lambda^2 F'' \left( R_0 \right)}{2}
\delta R
   - F' \left( R_0 \right) \Lambda^2 \cosh^2 x
\left( \delta \ddot\rho + 2 \delta \varphi'' - \delta \rho''
+ 2 \tanh x \delta\varphi' \right) \nn
& + 2 F' \left( R_0 \right) \Lambda^2 \delta \rho
+ F'' \left( R_0 \right) \Lambda^2 \cosh^2 x  \left( \tanh x \delta R'
+ \delta \ddot R \right) \, ,\\
\label{Nr8C}
0 = & - 2 \left( \delta{\dot\varphi}' + \tanh x \delta \dot\varphi \right)
+ \frac{F'' \left( R_0 \right) }{F' \left( R_0 \right) }\left(
\delta {\dot R}' + \tanh x \delta \dot R\right)\, ,\\
\label{Nr8D}
0 = & - \frac{ F'\left( R_0 \right) + 2 {\Lambda'}^2 F'' \left( R_0 \right)}{2}
\delta R
   - 2 {\Lambda'}^2 F' \left( R_0 \right) \delta\varphi
   - F' \left( R_0 \right)  \Lambda^2\cosh^2 x \left( - \delta \ddot\varphi +
\delta
\varphi'' \right) \nn
& - F'' \left( R_0 \right) \Lambda^2 \cosh^2 x \left( - \delta \ddot R + \delta
R'' \right) \, .
\end{align}
By combining (\ref{Nr8A}) and (\ref{Nr8B}) or by using (\ref{Nr8C}), we find
\be
\label{E6}
0 = -2 F'\left( R_0 \right) \delta \varphi + F'' \left( R_0 \right) \delta R\, .
\ee
Using Eqs.~(\ref{Nr7}), (\ref{Nr8A}), (\ref{Nr8B}), (\ref{Nr8C}),
(\ref{Nr8D}), we obtain
\begin{align}
\label{E7}
0 = & - \left( \frac{F'\left(R_0\right)}{F''\left(R_0\right)}
+ 4 {\Lambda'}^2 \right) \delta \varphi - 3\Lambda^2 \cosh^2 x \Box \delta
\varphi\, , \\
\label{E8}
0 = & - 2 \Lambda^2 \delta \rho + \left( 2 \Lambda^2 + \frac{3}{2}{\Lambda'}^2
- \frac{F'\left( R_0 \right)}{F''\left( R_0 \right)} \right) \delta \varphi \, .
\end{align}
We now assume (\ref{BH20}).
Because
\be
\label{E9}
\Box \delta\varphi = \left\{ - \omega^2 + \beta^2
+ \beta \left( \beta - 1 \right) \cosh^{-2} x \right\} \delta\varphi\, ,
\ee
we find
\be
\label{E10}
0 = - \omega^2 + \beta^2 \, , \quad
0= - \left( \frac{F'\left(R_0\right)}{F''\left(R_0\right)}
+ 4 {\Lambda'}^2 \right) - 3\Lambda^2 \beta \left( \beta - 1 \right)\, ,
\ee
which gives
\be
\label{E11}
\beta = \beta_\pm \equiv \frac{1}{2} \left\{ 1 \pm
\sqrt{ 1 - \frac{4}{3 \Lambda^2} \left(
\frac{F'\left(R_0\right)}{F''\left(R_0\right)}
+ 4 {\Lambda'}^2 \right) } \right\}\, .
\ee
Because there is always a solution where the real part of $\beta$ is positive,
the solution is always unstable. Furthermore if
\be
\label{E12}
1 - \frac{4}{3 \Lambda^2} \left( \frac{F'\left(R_0\right)}{F''\left(R_0\right)}
+ 4 {\Lambda'}^2 \right) < 0\, ,
\ee
the instability is accompanied with the vibration.
Thus, we described instability of RN black hole in Maxwell-$F(R)$ theory.

The radius of the horizon is given by (\ref{Nr21}),
if $\varphi_0 < 0$, $r_\mathrm{h}$ increases and there occurs the
anti-evaporation of RN black hole.
When $\beta$ and $\omega$ are complex,
instead of $\delta\varphi = \varphi_0 \cosh \omega t \cosh^\beta x$,
we obtain the following solution for $\varphi$
\be
\label{Nr26}
\delta\varphi = \Re \left\{ \left( C_+ \e^{\beta t} + C_+ \e^{- \beta t}
\right) \e^{\beta x}\right\} \, .
\ee
Here $C_\pm$ are a complex numbers. The notation $\Re$ expresses the real part.
As the real part of $\beta$ is always positive, $\delta\varphi$ increases
unless $C_+=0$ when $t$ increases and therefore the perturbation grows up,
which demonstrates that
the solution corresponding to the Nariai-like space-time is unstable.

Thus, we have shown that there could occur anti-evaporation even for the
extremal RN  ((anti-)de Sitter) black hole on the classical level
in $F(R)$ gravity. We should note that in the Einstein gravity, where
$F''(R)=0$,
there does not occur the anti-evaporation. This can be observed, for example,
in (\ref{E7}) and (\ref{E8}). When $F''(R)=0$, Eqs.~(\ref{E7}) and (\ref{E8}) show
that $\delta\varphi=\delta\rho=0$ and therefore there are no instabilities. 
We should note again that if we include the quantum effects, there could also be 
the usual Hawking radiation and evaporation. 

When $R_0>0$, there also appears the cosmological horizon. In case without charge, 
the anti-evaporation has been discussed in \cite{Nojiri:2013su}. 
We may expect that the qualitative structure of the anti-evaporation cold not been 
changed even if the black hole has a charge. 
It could be interesting if we consider the limit that all of the radii of the
three horizons coincide with each other. In this limit, there might appear the Nariai 
space-time in addition to the Nariai-like space-time in (\ref{BH18}). 
It could be interesting to investigate if the conditions of the anti-evaporation 
could be consistent between the Nariai space-time and the Nariai-like space-time. 

We should note that a new coordinate $x$ defined by (\ref{Nr3}) only covers 
the region between $r_0$ and $r_1$. 
In the second paper in \cite{Nojiri:2013su},  it has been shown that the 
perturbations in \cite{Nojiri:2013su} as in this paper are correct even if you 
consider most general coordinate system, which also describes the region outside 
the region $r_0\leq r \leq r_1$, for perturbations. 


\section{Discussion.}

In summary, we investigated the instabilities of RN black hole induced by
$F(R)$ gravity in the absence of Maxwell sector. It turns out that such
black holes may be the solution of $F(R)$ theory but only in the regions
with very strong effective gravitational coupling. Nevertheless, 
for inhomogeneous evolution of the universe the regions which may induce 
such charged black holes seem to be possible.
We also investigated the instabilities of RN black holes in Maxwell-$F(R)$
theory \cite{D}. It is demonstrated that the anti-evaporation effect is quite
realistic in such theory, as there are no strong limits on the value of
the effective gravitational coupling.
In relation with this, one can expect that strong magnetic fields at the
early universe are also somehow produced in the unified Maxwell-$F(R)$ theory
with significant contribution of $F(R)$ sector.
Furthermore, general (one or multi-horizon) spherically-symmetric solution is constructed  in
the Einstein frame of $F(R)$ gravity.

It is interesting that there is certain analogy between instabilities of
black holes in $F(R)$ gravity and those in GR with the account of
conformal anomaly \cite{brevik}. This is not strange because conformal
anomaly also includes higher derivative $R^2$-term (like some models of
$F(R)$) plus higher-derivative non-local terms. In fact, the first observation of the
anti-evaporation is done in GR with the account of conformal anomaly.

\appendix

\section{Connections and curvatures in the space-time (\ref{Nr3})}

For the space-time given in (\ref{Nr3}), the connections and the curvatures are
given by
\begin{align}
\label{Nr4}
& \Gamma^t_{tt}=\Gamma^t_{xx} = \Gamma^x_{tx} = \Gamma^x_{xt} = \dot\rho\, ,
\quad \Gamma^x_{xx} = \Gamma^x_{tt} = \Gamma^t_{tx} = \Gamma^t_{xt} = \rho'\, , \nn
& \Gamma^t_{\phi\phi} = \Gamma^t_{\theta\theta} \sin^2 \theta
= - \frac{\Lambda^2}{{\Lambda'}^2}
\dot\varphi \e^{-\left(\rho+\varphi\right)} \sin^2 \theta\, ,\quad
\Gamma^x_{\phi\phi} = \Gamma^x_{\theta\theta} \sin^2 \theta
= \frac{\Lambda^2}{{\Lambda'}^2}
\varphi' \e^{-\left(\rho+\varphi\right)} \sin^2 \theta\, ,\nn
& \Gamma^\theta_{t\theta} = \Gamma^\theta_{\theta t}
= \Gamma^\phi_{t\phi} = \Gamma^\phi_{\phi t} = - \dot \varphi\, ,\quad
\Gamma^\theta_{x\theta} = \Gamma^\theta_{\theta x}
= \Gamma^\phi_{x\phi} = \Gamma^\phi_{\phi x} = - \varphi'\, ,\nn
& \Gamma^\theta_{\phi\phi} = - \sin\theta \cos\theta\, ,\quad
\Gamma^\phi_{\phi\theta} = \Gamma^\phi_{\theta\phi} = \cot\theta\, ,\nn
& R_{tt} = - \ddot\rho + 2 \ddot\varphi + \rho'' - 2 {\dot\varphi}^2
   - 2\dot\rho \dot\varphi - 2 \rho' \varphi'\, ,\quad
R_{xx} = - \rho'' + \ddot\rho + 2 \varphi'' - 2 {\varphi'}^2 - 2 \dot\rho
\dot\varphi
   - 2 \rho' \varphi' \, ,\nn
& R_{tx} = R_{xt} = 2\dot\varphi' - 2 \varphi' \dot\varphi - 2 \rho'
\dot\varphi -2 \dot \rho \varphi'\, , \nn
& R_{\phi\phi} = R_{\theta\theta} \sin^2 \theta
= \left\{ 1 + \frac{\Lambda^2}{{\Lambda'}^2}
\e^{- 2 \left(\rho + \varphi\right)}\left( - \ddot\varphi + \varphi''
+ 2 {\dot\varphi}^2 - 2 {\varphi'}^2 \right) \right\} \sin^2\theta\, , \nn
& R = \Lambda^2 \left( 2\ddot\rho - 2 \rho'' - 4 \ddot\varphi + 4 \varphi''
+ 6 {\dot\varphi}^2  - 6 {\varphi'}^2 \right) \e^{-2\rho}
+ 2 {\Lambda'}^2\e^{2\varphi} \, ,\quad
\mbox{other components} = 0\, .
\end{align}
Here $\dot\ \equiv \partial/\partial t$ and $\ '\equiv \partial/\partial x$.

\section{General spherically symmetric solutions in $F(R)$ gravity}

In this work we consider a special class of black hole solutions, where the
scalar curvature is constant and the metric has a form given in (\ref{BH2}).
In this Appendix, we consider more general case, where the scalar curvature is
not constant and the metric is given in the general form  (\ref{BH1}) without
assuming (\ref{BH2}).

We now work in the scalar-tensor frame. The scalar field corresponds to the
scalar curvature.
Therefore as long as the scalar curvature is not singular, the scalar field is
not singular,either.
The scalar-tensor frame can be obtained by using the scale transformation and
therefore the causal structure of the space-time is not changed.

We now review the scalar-tensor description of the usual $F(R)$ gravity.
In $F(R)$ gravity, the scalar curvature $R$ in the Einstein-Hilbert action
\be
\label{JGRG6}
S_\mathrm{EH}=\int d^4 x \sqrt{-g} \left( \frac{R}{2\kappa^2} +
\mathcal{L}_\mathrm{matter} \right)\, ,
\ee
is replaced by an appropriate function of the scalar curvature:
\be
\label{JGRG7}
S_{F(R)}= \int d^4 x \sqrt{-g} \left( \frac{F(R)}{2\kappa^2} +
\mathcal{L}_\mathrm{matter} \right)\, .
\ee
One can also rewrite $F(R)$ gravity in the scalar-tensor frame.
By introducing the auxiliary field $A$, the action (\ref{JGRG7}) of the $F(R)$
gravity is rewritten in the following form:
\be
\label{JGRG21}
S=\frac{1}{2\kappa^2}\int d^4 x \sqrt{-g} \left\{F'(A)\left(R-A\right)
+ F(A)\right\}\, .
\ee
By the variation of $A$, one obtains $A=R$. Substituting $A=R$ into the action
(\ref{JGRG21}), one can reproduce the action in (\ref{JGRG7}).
Furthermore, we rescale the metric in the following way (conformal
transformation):
\be
\label{JGRG22}
g_{\mu\nu}\to \e^\vartheta g_{\mu\nu}\, ,\quad \vartheta = -\ln F'(A)\, .
\ee
Thus, the Einstein frame action is obtained as follows,
\begin{align}
\label{JGRG23}
S_E =& \frac{1}{2\kappa^2}\int d^4 x \sqrt{-g} \left( R -
\frac{3}{2}g^{\rho\sigma}
\partial_\rho \vartheta \partial_\sigma \vartheta - V(\vartheta)\right) \, ,\nn
V(\vartheta) =& \e^\vartheta g\left(
\e^{-\vartheta}\right) - \e^{2\vartheta}
f\left(g\left(\e^{-\vartheta}\right)\right)
= \frac{A}{F'(A)} - \frac{F(A)}{F'(A)^2}\, .
\end{align}
Here $g\left(\e^{-\vartheta}\right)$ is given by solving the equation
$\vartheta = - \ln F'(A)$ as $A=g\left(\e^{-\vartheta}\right)$.
Due to the scale transformation (\ref{JGRG22}),
the scalar field $\vartheta$ couples with the usual matter.

We now define a new scalar field $\phi$ by
\be
\label{FRBH1}
\sqrt{\omega(\phi)} d\phi = \sqrt{3}d\vartheta\, .
\ee
Here $\omega(\phi)$ can be an arbitrary function of the scalar field $\phi$.
Then the action (\ref{JGRG23}) can be rewritten as
\begin{align}
\label{FRBH2}
S_E =& \frac{1}{2\kappa^2}\int d^4 x \sqrt{-g} \left( R -
\frac{1}{2}\omega(\phi) g^{\rho\sigma}
\partial_\rho \phi \partial_\sigma \phi - U(\phi)\right) \, ,\nn
U(\phi) \equiv& V\left(\vartheta\left(\phi\right)\right) \, .
\end{align}
We now rewrite $A$ and $B$ in (\ref{BH1}) as $A=\e^{2\nu(r)}$ and
$B=\e^{2\lambda(r)}$.
We also assume that the scalar field only depends on the radial coordinate $r$,
$\phi=\phi(r)$.
Then further redefining the scalar field and using the ambiguity in the form of
$\omega(r)$, we may identify $\phi$ with $r$, $\phi(r)=r$.
Then the Einstein equations have the following form:
\begin{align}
\label{FRBH4}
\frac{2\nu'}{r} - \frac{\e^{2\lambda} - 1}{r^2}
=& \frac{1}{2} \omega(\phi) \left( \phi \right)^2 - \e^{2\lambda} V(\phi)\, ,\\
\label{FRBH5}
\frac{2\lambda'}{r} + \frac{\e^{2\lambda} - 1}{r^2}
=& \frac{1}{2} \omega(\phi) \left( \phi \right)^2 + \e^{2\lambda} V(\phi)\, ,\\
\label{FRBH6}
\nu'' + \left( \nu' - \lambda' \right)\nu' + \frac{\nu' - \lambda'}{r}
=& \frac{1}{2} \omega(\phi) \left(\phi \right)^2 + \e^{2\lambda} V(\phi)\, .
\end{align}
By using (\ref{FRBH5}) and (\ref{FRBH7})
\be
\label{FRBH7}
\e^{-2\lambda(r)}
= - \left( - \frac{1}{2r} + \frac{\nu'}{2} \right)^{-2}
\frac{\e^{-2\nu(r)}}{r^4}
\int dr r^2 \e^{2\nu(r)} \left( - \frac{1}{2r} + \frac{\nu'}{2} \right)\, .
\ee
Then if we give the form of $\nu(r)$, we obtain the form of $\lambda(r)$
up to a constant of the integration.
Furthermore by using (\ref{FRBH4}) and (\ref{FRBH5}), one finds
\be
\label{FRBH8}
\omega(r) = \frac{2 \left( \nu'(r) + \lambda'(r)\right) }{r}\, , \quad
U(r) = 2 \left\{ \frac{- \nu'(r) + \lambda'(r)}{r}
+ \frac{\e^{2\lambda(r) - 1}}{r^2} \right\}\e^{-2\lambda(r)}\, .
\ee
Then if we give an arbitrary but explicit form of $\nu$, we can determine
the form of $\lambda$ by (\ref{FRBH7}) and those of $\omega$ and $U$.

For example, as in case of the Schwarzschild black hole, where
\be
\label{FRBH9}
\e^{2\nu(r)} = 1 - \frac{r_0}{r}\, ,
\ee
with a constant $r_0$, Eq.~(\ref{FRBH7}) gives
\be
\label{FRBH10}
\e^{-2\lambda(r)}
= - \left( - \frac{r}{2} + \frac{3r_0}{4} \right)^{-2}
\left(  1 - \frac{r_0}{r} \right) \left( - \frac{r^2}{4}
+ \frac{3r_0 r}{4} + C \right)\, .
\ee
Here $C$ is a constant of the integration. If we choose,
\be
\label{FRBH11}
C=\frac{9r_0^2}{16}\, ,
\ee
we surely obtain the Scwarzschild solution:
\be
\label{FRB12}
\e^{-2\lambda(r)}
=\e^{2\nu(r)} = 1 - \frac{r_0}{r}\, .
\ee
For the Schwarzschild solution (\ref{FRB12}), by using (\ref{FRBH8}),
we find $\omega=U=0$,  because the Schwarzschild solution is the
vacuum solution of the Einstein equation.

We may consider the case that there are two horizons
\be
\label{FRBH13}
\e^{2\nu(r)} = \left( 1 - \frac{r_1}{r} \right) \left( 1 - \frac{r_2}{r}
\right) \, ,
\ee
with constants $r_1$ and $r_2$ which correspond to the horizon radiuses.
Then we obtain
\be
\label{FRBH14}
\e^{-2\lambda(r)}
= \left( 1 - \frac{r_1}{r} \right) \left( 1 - \frac{r_2}{r} \right)
\left( - \frac{r}{2} + \frac{3\left( r_1 + r_2 \right)}{4} - \frac{r_1 r_2}{r}
\right)^{-2}
\left( - \frac{r^2}{4} + \frac{3}{4} \left(r_1 + r_2\right) r
   - r_1 r_2 \ln \frac{r}{r_0} \right)\, .
\ee
Here $r_0$ is a constant of the integration.
Because
\be
\label{FRBH15}
   - \frac{r}{2} + \frac{3\left( r_1 + r_2 \right)}{4} - \frac{r_1 r_2}{r}
= - \frac{1}{2r}\left[
\left( r - \frac{3}{4} \left(r_1 + r_2\right) \right)^2 + 2r_1 r_2
   - \frac{9}{16} \left( r_1 + r_2 \right)^2 \right]\, ,
\ee
we find $\e^{-2\lambda}$ is not singular anywhere if
\be
\label{FRBH16}
2r_1 r_2 > - \frac{9}{16} \left( r_1 + r_2 \right)^2 \, .
\ee
On the other hand, because
\be
\label{FRBH17}
- \frac{r^2}{4} + \frac{3}{4} \left(r_1 + r_2\right) r
   - r_1 r_2 \ln \frac{r}{r_0}
= - \frac{1}{4} \left( r - \frac{3\left(r_1 + r_2\right)}{2} \right)^2
+ \frac{9}{16} \left( r_1 + r_2 \right)^2  - r_1 r_2 \ln \frac{r}{r_0}
< r_1 r_2 \left( 2 - \ln \frac{r}{r_0} \right)\, ,
\ee
we find $\e^{-2\lambda}$ vanishes at $r=\e^2 r_0$ in addition to  the horizons
$r=r_1$, $r_2$. We may choose $r_0$ to be small enough so that the surface
$r=r_0$ becomes deep inside the horizons.
Near $r=r_0$, the spacial part in the space-time has a form of throat.
By using (\ref{FRBH8}), we find
\begin{align}
\label{FRBH18}
\omega(r)
=& \frac{r}{2}
\left( - \frac{r}{2} + \frac{3\left( r_1 + r_2 \right)}{4} - \frac{r_1
r_2}{r}\right)^{-1}
\left( - \frac{r^2}{4} + \frac{3\left(r_1 + r_2\right) r}{4}
   - r_1 r_2 \ln \frac{r}{r_0} \right)^{-1} \nn
& \times \left\{ \left(- \frac{1}{2} + \frac{r_1r_2}{r^2}\right)
\left( - \frac{r^2}{4} + \frac{3}{4} \left(r_1 + r_2\right) r
   - r_1 r_2 \ln \frac{r}{r_0} \right) 
   - \frac{1}{2}
\left( - \frac{r}{2} + \frac{3\left( r_1 + r_2 \right)}{4} - \frac{r_1
r_2}{r}\right)^2
\right\} \, , \\
\label{FRBH19}
V(r) =& \frac{2}{r^2} - \frac{r}{2} \left( \frac{r_1 + r_2}{r^2} - \frac{2 r_1
r_2}{r^3} \right)
\left( - \frac{r}{2} + \frac{3\left( r_1 + r_2 \right)}{4} - \frac{r_1
r_2}{r}\right)^{-2}
\left( - \frac{r^2}{4} + \frac{3}{4} \left(r_1 + r_2\right) r
   - r_1 r_2 \ln \frac{r}{r_0} \right) \nn
& + \frac{2}{r} \left( 1 - \frac{r_1}{r} \right) \left( 1 - \frac{r_2}{r}
\right)
\left( - \frac{r}{2} + \frac{3\left( r_1 + r_2 \right)}{4} - \frac{r_1
r_2}{r}\right)^{-3} \nn
&\times \left\{ \left(- \frac{1}{2} + \frac{r_1r_2}{r^2}\right)
\left( - \frac{r^2}{4} + \frac{3}{4} \left(r_1 + r_2\right) r
   - r_1 r_2 \ln \frac{r}{r_0} \right)
   - \frac{1}{2} \left( - \frac{r}{2} + \frac{3\left( r_1 + r_2 \right)}{4}
   - \frac{r_1 r_2}{r}\right)^2 \right\}\, .
\end{align}
Therefore both of $\omega$ and $U$ are not singular nor vanish
at the horizon $r=r_1$, $r_2$.
Similarly, we may consider multi-horizon solutions by assuming
\be
\label{FRBH20}
\e^{2\nu(r)} = \prod_{i=1}^n \left( 1 - \frac{r_i}{r} \right)\, .
\ee
Thus, we found general spherically-symmetric solutions in $F(R)$ gravity
in the Einstein frame. The instabilities of any specific multi-horizon black
hole  from this class may be studied in the same way as in this work.

\section*{Acknowledgments.}

We are grateful to S. Sushkov for helpful discussion.
The work by SN is supported in part by Global COE Program of Nagoya University
(G07) provided by the Ministry of Education, Culture, Sports, Science \&
Technology and by the JSPS Grant-in-Aid for Scientific Research (S) \# 22224003
and (C) \# 23540296 and that by SDO is supported in part by  MINECO 
(Spain), project FIS2010-15640 and by  grant of 
Russ. Min. of Education and Science, project TSPU-139.

\end{document}